\documentclass{article}

\usepackage{arxiv}

\usepackage[utf8]{inputenc} % allow utf-8 input
\usepackage[T1]{fontenc}    % use 8-bit T1 fonts
\usepackage{hyperref}       % hyperlinks
\usepackage{url}            % simple URL typesetting
\usepackage{booktabs}       % professional-quality tables
\usepackage{amsfonts}       % blackboard math symbols
\usepackage{nicefrac}       % compact symbols for 1/2, etc.
\usepackage{microtype}      % microtypography
\usepackage{lipsum}

\title{A Noxious Market for Personal Data}

\author{
  Abdul Z. Abdulrahim\thanks{Abdul Abdulrahim is also a Graduate Student in the Department of Computer Science, University of Oxford, Oxford, United Kingdom (e-mail: abdul.abdulrahim@cs.ox.ac.uk).},   Michael Famoroti \\
  Stears\\
  Lagos, Nigeria \\
  \texttt{[abdul.abdulrahim, michael.famoroti]@stearsng.com}
}

\begin{document}
\maketitle

\begin{abstract}
Many policymakers, academics and governments have advocated for exchangeable property rights over information as it presents a market solution to what could be considered a market failure. Particularly in jurisdictions such as Africa, Asia or South America, where weaker legal protections and fleeting regulatory enforcement leaves data subjects vulnerable or exploited regardless of the outcome. We argue that whether we could achieve this personal data economy in which individuals have ownership rights akin to property rights over their data should be approached with caution as a solution to ensuring individuals have agency over their data across different legal landscapes. We present an objection to the use of property rights, a market solution, due to the \textit{noxious} nature of personal data --- which is founded on Satz \cite{satz2010some} and Sandell \cite{sandell1998money}'s objection to markets. 
% Ultimately, our rights over personal data and privacy are borne out of our basic human rights and are a precondition for the self-development, personal fulfilment and the free enjoyment of other fundamental human rights --- and putting it up for sale risks corrupting its essence and value. 
\end{abstract}

% keywords can be removed
% \keywords{First keyword \and Second keyword \and More}

\section{Introduction}
% \includecomment{Presented at NeurIPS 2019 Workshop on Machine Learning for the Developing World.}

McNealy's statement; ``Privacy is dead -- get over it'' in 2000 struck a chord with many spectators as it highlighted the weaknesses in our privacy frameworks with the advent of government surveillance and data-mining. It forced some legal scholars to ask what we could learn from copyright law for ownership rights over personal data and privacy to mitigate the risks from the use of data by large technology companies. Some have advocated for exchangeable property rights over information as it presents a market solution to what could be considered a market failure. While we concede that there are issues with our current privacy frameworks and the motivations for enhancing privacy rights, we see this approach as fundamentally problematic. We argue that whether we could achieve this \textit{personal data economy} in which individuals have ownership rights akin to property rights should be approached with caution as our focus should be to protect the essence of privacy. 

% \section{Enhancing Privacy: Incompetence or Intractable?}
Privacy in its simplest form is ``the right to be let alone, or freedom from interference or intrusion'' \cite{IAPP, wachter2017privacy}. In the context of data and information, it can be interpreted as the right to have control over the use of our personal information \cite{IAPP}. While there are many other variations of how privacy could be defined, we will avoid this philosophical debate which requires a separate discussion. It is also important to highlight that since Samuelson's \cite{samuelson1999privacy} paper evaluating the extension of intellectual property rights to personal data, some privacy frameworks have changed significantly, especially in Europe. With the implementation of the General Data Protection Regulation (GDPR) and more regulation in the pipeline, the scope for data protection has expanded. 

% Thus, while some of the arguments posed on the weaknesses of our framework are still present, some countries have made some good strides in increasing personal data protection.

That being said, it is still remarkably inexpensive and easy for large companies to collect substantial amounts of information identifiable to particular individuals. Even though our personal data may be purported to be anonymised, Ohm \cite{ohm2009broken} and Schneier's \cite{schneier2015data} account of the failure of anonymisation and the competing interests between security and privacy, casts doubts on the motivations and competence of governments in emerging markets to protect our privacy rights and our personal data appropriately. Furthermore, Wachter's \cite{wachter2018right} work accounting for how partial information derived from personal data (using \textit{inferential analytics}) can still provide just as useful privacy-compromising insights demonstrate the intractability of the problem. As such, it is not just our motivations that need correcting, but technological advancements also subvert attempts to enhance privacy rights.

Given the incentives for firms to: (i) collect and process personal data to improve their services; (ii) amass personal information as an asset that gives them a competitive advantage; and (iii) use information to create market barriers to entry, it comes as no surprise that this has been characterised as a market failure. One, some would argue, is best solved using a market approach that gives tradable rights over personal data to individuals.
% (c) Still seems to favour institutions, e.g. Trade Secrets Directive (check for US examples)

\section{Property Rights: A Way Forward}
Lessig \cite{lessig1999code} argues that the future shape of the Internet depends on the actions we take to define it. Interestingly, in his discussion of privacy, he ultimately settles on a market-based allocation of privacy interests that allows for trading information. Lessig's \cite{lessig1999code} proposition for enhanced privacy protection is no doubt supported by those who conceive of personal data protection as a fundamental civil liberty interest \cite{davies1997re}, those who suggest cognitive limitations on the ability of individuals to comprehend and accurately assess the risks of revealing personal data \cite{privacy1998washpost} and those who argue for information privacy protection to guard against identity theft, harassment, and other wrongful uses of personal information \cite{kang1997information}. In fact, the market based solution suggested by Lessig \cite{lessig1999code} continues to grow in support (see \cite{murphy2017property}, \cite{bartlett1999developments} and \cite{mell1996seeking}). 

The extension of property rights to personal data perhaps satisfies this need for better individual control, primarily when these rights can be traded. Key to this line of thinking is that owning our personal data gives us the freedom to choose. The freedom to choose to engage or disengage with the market, to be remunerated for the use of our data and with whom, when and where we share our personal data. However, this freedom comes with risks and practical limits on enforcing this choice. 

Rotenberg \cite{rotenberg2001fair} and Samuelson \cite{samuelson1999privacy} present a strong case for why the use of tradable property rights is flawed using arguments against its practicality, complexity and morality. However, none put forward a comprehensive case for the fundamental objection to the use of property rights and markets as a solution. In the next section, we argue that the idea of extending property rights to privacy rights creates a noxious market. It inches us closer to a not-too-far dystopian future where we essentially pay for our privacy in what is termed the ``\textit{personal data economy}'' by Elvy \cite{elvy2017paying}. Hence, while it may economically and technically make sense to have tradable property rights, not every good or service was made to be sold. 

\section{Noxious Markets: Preying on the Weak}

Rotenberg \cite{rotenberg2001fair} and Samuelson \cite{samuelson1999privacy} made some valid criticisms of the use of a market's approach to solving the issues of enforcing privacy rights. More recently, Elvy \cite{elvy2017paying} also outlined how the transformation of privacy into a tradable product may worsen unequal access to privacy and enable predatory and discriminatory behaviour. Elvy's \cite{elvy2017paying} criticisms are borne out of a fundamental objection to the use of markets to organise and allocate certain goods and services (see \cite{satz2010some, sandell1998money}). We summarise these as follows:

\begin{itemize}
    \item[(a)] A good should not be sold because you should not or cannot even own it, e.g., ownership in another person.
    \item[(b)] A good should not be sold because you should not be able to give up your claim to that good, e.g., a right to vote. 
    \item[(c)] A good should not or cannot be sold because doing so may have: (i) adverse effects on the good and its provision; (ii) adverse effects on the buyer, the seller and their relationship, e.g., friendship.
\end{itemize}

At least in the European context, Article 8 of the Charter of Fundamental Rights of the European Union gives European residents some ownership of their personal data and how it is used --- though not tradable, --- so the first objection may not technically apply in this case. As for the second, we could argue that our right to privacy is akin to suffrage (the right to vote); necessary for effective democracy and enabling other human rights. The issue, however, is that privacy is not a universal concept and it is not clear that the right to privacy and our personal data is on the same level as our suffrage or other human rights. While privacy is arguably necessary for preventing the exploitation of individual autonomy by competing interests and enabling other fundamental human rights, it is a right that is still developing, and its inalienability is still in question \cite{wachter2017privacy}. Due to the limited scope of this paper, we do not focus on this objection as it brings up the more significant point of the relative normativity of legal principles and rights \cite{meron1986hierarchy}.

The third objection is also at the heart of why privacy should not be a property right. Satz \cite{satz2010some} builds a case on this objection for why goods or services that result in \textit{noxious markets} should not be commoditised. Deriving from her arguments, the logic is; because markets affect our relationship with others, \textit{noxious markets} raise problems for the standing or relationship of the parties, before, during and after the process of market exchange \cite{satz2010some}. The problem herein lies not with the meaning of goods or services, but with the fact that some markets undermine our standing as equals. Consequently, commoditising our personal data (and in turn privacy rights) creates a \textit{noxious market} that threatens the requirements of equal standing captured in the ideal of democratic citizenship --- that is, equal fundamental political rights and freedoms, and equal rights to a fair share of social welfare and equality of opportunity. In the context of privacy rights, this results from the weak or asymmetric knowledge or agency we have as individuals to comprehend the risks or potentials from the use of our personal data and our underlying vulnerability against large institutions \cite{elvy2017paying, satz2010some}. It is evident that in a personal data economy the balance of power does not favour the individual, and to assume that giving us agency to negotiate our tradable rights against large institutions in any way enhances our privacy is naive. The likely result of such commodification is an individually adverse outcome which violates our welfare and agency interests, and a collectively harmful outcome to markets in society as a scheme of cooperation as equals. 

\section{The Reality of ``Stomach Infrastructure''}
The problem with a market for personal data is the arguably unfavourable equilibrium that results because individuals are unable to properly value their personal data. The benefits of keeping personal data private may not be immediate and can spread out into the future. As people have a present-bias/time-inconsistent preferences \cite{hoch1991time, thaler2016behavioral}, they are liable to value the privacy of their data much lower than their future selves would, particularly at the time when data privacy is most important. For example, shoppers may be happy to release personal data like home address and occupation as part of a loyalty program, but would not appreciate if this information was used to set tailored prices.\footnote{See Surge pricing comes to the supermarket, The Guardian UK, 4 June 2017} 

Estimating the value that can be gleaned from your personal data is almost an impossible calculation for an individual to make. This is particularly true as the value of the data depends on what can be layered or added to it, and that is information the individual is unlikely to have. Since this knowledge is more likely to rest with large institutions (governments, corporates, etc.), this group would have substantial market power to set prices. Furthermore, there may be negative externalities involved in a private transaction to sell personal data. Advances in inferential analytics means that institutions are able to infer useful information about individuals in a given group provided they have sufficient data on members of the same group \cite{wachter2018right}. Again, the external costs on others are likely to be ignored when an individual decides to transact in the market for private data.

The fact that individuals are unable to properly value their personal data is supported by the ``privacy paradox'', the phenomenon where people profess a desire to preserve their personal data yet often trade it away for free \cite{barnes2006privacy}. Nowhere is this more evident than when we look at the less well-off. The poor are much more likely to participate in the personal data market as they would have the lowest willingness-to-accept ranges in society. Although not constrained by individual wealth like willingness to pay (WTP), willingness to accept (WTA) is likely to be a function of wealth in a similar way \cite{hanemann1991willingness, horowitz2003willingness}. For example, a millionaire would likely place a higher price on revealing his home address than a college student. 

This has implications in circumstances where the value of the data is independent on income (or inversely related) such as voting behaviour or credit repayment history. Poorer individuals are more likely to be in situations where they feel they have to sell their personal data. In other words, the choices poor individuals make in a market society may not be entirely free and voluntary as they may feel compelled to make just to survive. A related example can be found in India’s blood farms, where poor migrants were relatively high sums to donate blood.\footnote{See Blood for sale: India's illegal 'red market', BBC News, 27 January 2015} Unfortunately, the methods used at the farms escalated and weak migrants were coerced into selling more blood than is deemed healthy.

The implication of all this is that creating a market for private data would essentially be creating a market for poorer people’s personal data. This dynamic is worsened by the potential abuses of personal data by institutions. The more data/information an institution has, the more it is able to influence an individual. Creating a market for personal data leaves the poor uniquely vulnerable to institutional influence in a way that can easily perpetuate existing inequalities.
This scenario is best depicted by looking at patterns of electoral malpractice in Nigeria. Election campaigns in Nigeria have long been characterised by the implicit exchange of gifts for community votes (termed ``stomach infrastructure''). These gifts could come in the form of small infrastructure projects, like a new transformer to serve the community, or non-durable consumer goods like bags of rice and palm oil. 

During recent sub-national elections in Ekiti State, a southwest state that is one of the poorest in the country, this implicit transaction was replaced by an explicit underground market for individual votes.\footnote{INEC decries ‘vote-buying’ in Ekiti governorship election, Punch Newspapers Nigeria, 18 July 2018} Two patterns emerged in this vote-buying behaviour. The first is that vote-buying was more prevalent in less developed communities further away from the state capital, and the second is that the market price for votes differed across communities. Both trends tally with the prediction that poorer individuals are more likely to partake in the market for ``commodities'' like votes and personal data. In a society as unequal as Nigeria, the ballot box is one of the few equalisers (``one person, one vote'') so the value of a vote is presumably high (even without considering its political value). A similar situation can be envisaged with a market for personal data where individuals undervalue their personal data and poorer individuals are more likely to sell their personal data.

\section{Conclusion}

If we are so inclined to put our personal data up for sale, policy responses could entail restricting the types of data a person may be able to give up under normal circumstances, i.e. requiring onerous procedures when consenting to the sale of sensitive personal data. Another option is guaranteeing perpetual rights to data subjects over data that has a long shelf-life or usefulness to controllers to ensure a subsisting interest in their favour. Although controversial in many ways, what these policy responses ensure is that data subjects are not exploited given the rapid advancement in subversive techniques, the lack of understanding of the true harm or cost to losing our privacy and the ability to reverse the harm. To put an individual back in the position he or she would have been in absent the harm caused by the misuse of their personal data or violation of their privacy rights --- their `original position' \cite{posner1979ethical}. Proponents of a property rights approach to privacy rights may argue that these arguments do not provide adequate correction to the market failure for privacy and data. Nonetheless, we believe their objections are founded in the lacklustre response of our governments and policymakers to keep up with the rapid development in technology and its subversive uses. 

% Libertarians may also have further objections to the paternalistic nature of these policies, but history has shown us that free markets are ineffective at dealing with pre-existing or systemic inequalities and vulnerabilities.

Our right to privacy emanates from our human rights, and we be careful not to corrupt that. We should look for frameworks that protect the essence of privacy and not exploit them. As it is currently regulated, the right to privacy is predominantly conceived as a subjective right protecting the individual interests from interference \cite{van2018new}. Although the existing `infringement'-criterion works well when applied to more traditional privacy violations, such as third-party eavesdropping on a private conversation, concerning modern data-driven technologies, it is often difficult to demonstrate an actual and concrete infringement, for example with \textit{inferential analytics}. Therefore, an increasing number of scholars, with whom we agree, advocate the use of the `non-domination' principle. The core of this principle is not whether there has been an ‘interference’ with a right. Instead, it looks at existing power relations and the potential for the abuse of power. The use of markets presents the potential for exacerbating pre-existing inequalities and unequal access to privacy. Granted, the motivations behind the push for property rights is grounded in a valid fear for weak privacy frameworks that can and should be enhanced. Nevertheless, not every good or service was meant to be traded, and our privacy rights should never be for sale --- even when it is masked behind the idea of privacy enhancement.

\bibliographystyle{unsrt}  
\bibliography{references}  %%% Remove comment to use the external .bib file (using bibtex).


\end{document}